\documentclass{article}
\usepackage{amssymb}

\input{tcilatex}

\begin{document}

\title{A Strong Constraint on Ever-Present Lambda }
\author{John D. Barrow \\
%EndAName
DAMTP, Centre for Mathematical Sciences,\\
Cambridge University, Wilberforce Rd.,\\
Cambridge CB3 0WA, UK}
\date{}
\maketitle

\begin{abstract}
We show that the causal set approach to creating an ever-present
cosmological 'constant' in the expanding universe is strongly constrained by
the isotropy of the microwave background. Fluctuations generated by
stochastic lambda generation which are consistent with COBE and WMAP
observations are far too small to dominate the expansion dynamics at $\
z<1000$ and so cannot explain the observed late-time acceleration of the
universe. We also discuss other observational constraints from the power
spectrum of galaxy clustering and show that the theoretical possibility of
ever-present lambda arises only in 3+1 dimensional space-times.

\end{abstract}

The apparent existence of a non-zero cosmological constant, $\Lambda ,$ with
a positive value of order $10^{-120}$ in Planck units is a mystery to
astronomers and a challenge to the ingenuity of theoretical physicists. It
may be explained by some unknown, or as yet partially known, fundamental
theory of everything which prescribes the vacuum state of the universe
uniquely and completely. Equally, it might arise through a random
symmetry-breaking process within a landscape containing a huge (or even an
infinite) number of different possible vacua. In the latter case, it would
make no more sense to try to predict the observed value of $\Lambda $ from
the underlying theory than to use a theory of dynamics to predict how many
planets there should be in the Solar System. All a theory of $\Lambda $
could do is to assess the likelihood of the observed value within the subset
of outcomes that permit the evolution of 'observers'. As first shown in ref 
\cite{bt}, the constraint on $\Lambda $ from the requirement that galaxy and
star formation be possible places an upper bound on the allowed magnitude of 
$\Lambda $ that we (or other 'observers') could observe that is only about
an order of magnitude weaker than the observed value. The observed value of $%
\Lambda $ \cite{sn} implies that the universe has accelerated for the last
25\% of its expansion scale-factor history, and the energy densities
contributed by $\Lambda $ and by the other material stresses in the
Friedmann equation governing the expansion of the universe are still of
comparable order. Thus, there are two puzzles about $\Lambda $: why is it
non-zero and why did it become dynamically significant so close to the
present epoch?

One specific attempt to address these questions with a simple testable model
is that proposed by Ahmed et al \cite{ahmed}, which develops an earlier
simpler proposal by Sorkin \cite{sork}. It notes that the magnitude of the
observed $\Lambda $ in Planck units is of order $N^{-1/2},$ where $N$ is the
spacetime volume of the universe (Hubble volume $\times $ expansion age $%
\propto t^{4}$) in Planck units. Thus, at comoving proper expansion time $t,$
we have $N(t)\sim (t/t_{p})^{4}\ $and $\Lambda \sim N^{-1/2}\sim
(t_{p}/t)^{2},$ where the Planck time is $t_{p}=(G\hbar /c^{5})^{1/2}\sim
10^{-43}s$. Therefore, today, at $t_{0}\sim 10^{17}s$, we should observe $%
\Lambda $ as a residual quantum gravity effect with a magnitude

\begin{equation}
\Lambda \sim (N(t_{0}))^{-1/2}\sim (t_{p}/t_{0})^{2}\sim 10^{-120}.
\label{lam}
\end{equation}

In this model the induced value of $\Lambda $ at any time $t$ will always be
of order $(t_{p}/t)^{2},$ and so is always of the same order as both the
dominant matter or radiation density and the square of the Hubble expansion
rate $H^{2}$ in the Friedmann equation. It is for this reason that a $%
\Lambda $ stress originating in this fashion has been called 'ever-present
lambda'.

We note that this type of 'ever-present' scenario is only a possibility in a
universe with three space dimensions. If the universe has $D$ spatial
dimensions then the spacetime volume grows as $t^{D+1},$ and the magnitude
of the lambda density induced by Poisson fluctuations in the number of
Planck-sized regions in the whole $(D+1)-$dimensional spacetime volume is $%
\rho _{\Lambda }\propto t^{-(D+1)/2}$. However, the Hubble parameter and the
dominant matter (or radiation) density still evolve at $H^{2}\sim \rho \sim
t^{-2}$ and so are only always of comparable order to $\rho _{\Lambda }$ for
worlds with $D=3$. If $D>3,$ then $\rho _{\Lambda }$ falls off faster than $%
t^{-2}$ and is never $O(\rho )$ at late times; whereas, if $D<3,$ $\rho
_{\Lambda }$ dominates $\rho $ at late times.

Ahmed et al \cite{ahmed} propose a specific mechanism by which the induced $%
\Lambda $ term can arise as a Poisson fluctuation in $N(t)$ by means of a
causal-set description of the geometry of four-dimensional spacetime, which
is reviewed in ref \cite{ahmed, causal}. Fluctuations arise as the number of
causally connected Hubble four-volume grows with the expansion of the
universe to encompass more Planck-sized spacetime volumes \cite{ahmed}. The
associated quantum uncertainty principle is $\delta \Lambda \delta N\gtrsim
1 $, with $\delta N\sim N^{1/2}$. Specifically, the induced $\Lambda $ is
assumed to arise from a Poisson process on the number of independent
Planck-sized spacetime volumes contained within the particle horizon at time 
$t$. Hence, in this discrete model of space-time, $N$ is equal to the number
of Planck-sized volumes in the total space-time volume, and so $N=V$. If at
the $n^{th}$ timestep, $t_{n}$, this number of Planck four-volumes is $N_{n}$%
, we can define its change with respect to the number of Planck-sized
spacetime volumes in the horizon at time $t_{n}$, given by $V(t_{n})$, by
the difference

\begin{equation}
\delta N_{n}\equiv N_{n+1}-N_{n}=V(t_{n+1})-V(t_{n}).  \label{f1}
\end{equation}%
Hence, at the $(n+1)^{st}$ timestep, the induced $\Lambda $ will be taken to
contribute an energy density to the Friedmann equation equal to

\begin{equation}
\rho _{\Lambda ,n+1}=\frac{S_{n+1}}{N_{n+1}}=\frac{S_{n}+\alpha \xi _{n+1}%
\sqrt{\delta N_{n}}}{N_{n}+\delta N_{n}}\ ,  \label{f2}
\end{equation}%
where $S_{i}$ are the sums of the first $i$ random numbers with $S_{i=0}=0$.
The Poisson process is assumed to generate random numbers $\xi _{i}$ with
zero mean and standard deviation $1$ and the fluctuation intensity is
controlled by the free constant $\alpha $. The contribution of\ the induced $%
\rho _{\Lambda \ \ }$to the Friedmann equation takes the usual form:

\begin{equation}
3H^{2}=\rho _{m}+\rho _{\gamma }+\rho _{_{\Lambda }},  \label{frw}
\end{equation}%
where $\rho _{m}$ and $\rho _{\gamma }$ are the densities of matter and
radiation. Numerical simulations \cite{ahmed} confirmed that the induced $%
\rho _{\Lambda \ \ }$term is always similar in magnitude to the largest of $%
\rho _{m}$ and $\rho _{\gamma }$ during the evolution of the universe if $%
\alpha $ is sufficiently large. There are some technical issues arising from
this model: it will have to be cut off at the extremes to stop very large
negative contributions to $\Lambda $ being inconsistent with the positivity
of the left-hand side of (\ref{frw}), (note that 50\% of the time the
induced lambda fluctuations are negative) and the quantum physics of the
causal-set generation of the fluctuations within the context of some
unimodular theory of gravity remains to be explored in detail \cite{causal}.
The occurrence of large negative fluctuations can be rendered innocuously
improbable over the life of the universe by choosing $\alpha $ to be
sufficiently small and we shall ignore these difficulties. The whole model
could be rendered more rigorous by adopting to a full stochastic
differential equation formulation. However, we do not intend to investigate
these aspects of the scenario here. Rather, we are interested in the best
observational bounds on the allowed value of the undetermined statistical
intensity parameter, $\alpha $, which is the one free parameter in the model.

In their study, Ahmed et al \cite{ahmed} consider three principal
constraints on $\alpha $: $\alpha $ must not be too small if we want the
universe to accelerate in the recent past because we need the induced lambda
fluctuations to permit $\rho _{\Lambda }\gtrsim \rho _{m}$ today, as
observed \cite{sn}, with reasonable probability; $\alpha $ must not be too
large, or large negative lambda fluctuations will be too probable and
contradict $H^{2}>0$ in eqn. (\ref{frw}); $\alpha $ must not be so large as
to create changes in the expansion rate of the universe at redshift $z\sim
10^{10}$ that lead to big bang nucleosynthesis of unacceptable abundances of
helium-4 with significant likelihood. There is some tension between these
opposing requirements but the best compromise appears to be for the range
with $\alpha =0.01-0.02$. It might also be the case that an alternative
formulation could require the lambda fluctuations to be positive
semi-definite.

There is another constraint on the scenario that might be expected to
produce an upper bound of $\alpha \lesssim 0.02$ or better$.$ The induced
lambda contributions are of order $\alpha $ times the dominant density that
drives the expansion. At cosmological times close to the epoch of
matter-radiation equality, at redshift $1+z_{eq}\sim 2.4\times 10^{4},$
there will be small corrections to the equal-density time in the
ever-present lambda model that can be determined accurately by a simulation.
However, they must be smaller than about $2\%$ in amplitude or they will
shift the epoch of equal density sufficiently to move the peak of the power
spectrum of galaxy clustering away from its observed position with
significant probability. This is a similar bound to that used by Tegmark 
\cite{teg} to constrain light neutrino masses, and Liddle et al \cite{lid}
to constrain Brans-Dicke theories. It arises because density perturbations
in the matter only commence growth by gravitational instability after the
radiation-dominated era ends.

In the earlier analysis of ref. \cite{ahmed}, it was assumed that the
stochastic process generating the lambda fluctuations is perfectly
homogeneous and isotropic, so $\rho _{\Lambda }\equiv \rho _{\Lambda }(t)$.
Thus, in effect, it simply transforms the dynamics along the line of exact
Friedmann universes containing matter, radiation and instantaneous
lambda-like stresses. We believe that this is not a physically complete
representation of the underlying stochastic process. Rather, we would expect
the fluctuation to be at best only statistically homogeneous and isotropic
because of its quantum gravitational origin. As a result of quantum
uncertainty, there will always be small statistical fluctuations in the
overall homogeneity and isotropy of the expansion dynamics on the horizon
scale, yielding $\rho _{\Lambda }\equiv \rho _{\Lambda }(\mathbf{x},t)$.
These will manifest themselves as small gravitational potential fluctuations
on the horizon scale. When the universe cools sufficiently for atomic
recombination to occur at the epoch of last scattering of the microwave
background photons, at $z_{ls}\sim 1000$, the horizon-sized stochastic
fluctuations will create temperature anisotropies in the microwave
background of amplitude $\alpha .$ They will be equal in amplitude to the
gravitational potential fluctuations produced by the density perturbations
on a scale $L\sim t$, across angular scales of order a few degrees. These
fluctuations contribute an amplitude to the observed microwave temperature
anisotropy, $\Delta T/T$, that is a fraction $\alpha $ of the background
density:

\begin{equation}
\Delta T/T\sim \Delta \Phi /\Phi \sim (\delta \rho _{_{\Lambda }}/(\rho
+\rho _{_{\Lambda }}))(L/t)^{2}\ \sim \alpha ,  \label{cmb}
\end{equation}%
where $\Delta \Phi $ is the gravitational potential perturbation, with $%
\Delta \Phi \sim \delta \rho _{_{\Lambda }}L^{2}$ from the Poisson equation,
and $\Delta T$ is the angular anisotropy perturbation observed in the cosmic
microwave background temperature. The observational data from the COBE \cite%
{cobe} and WMAP \cite{wmap} missions, along with complementary ground-based
detections, therefore provide the powerful constraint:

\begin{equation}
\alpha \lesssim 10^{-5}.  \label{a}
\end{equation}%
This an extremely tight bound on the causal set mechanism for ever-present
lambda generation and prevents any effective lambda generated by this
mechanism from ever being large enough to dominate the dynamics at $0\leq
z<1000$ with any significant probability, as is required to provide an
explanation for the redshift-distance relation of type Ia supernovae at low
redshifts \cite{sn}. This would require a more complicated stochastic
generation model with an effective $\alpha $ that increased with time so
that it could be $O(1)$ near the present but $O(10^{-5})$ at $t_{rec}\sim
10^{13}s$. This type of limit should also severely constrain other proposals
for the generation of small$\ \Lambda $ values by means of stochastic
space-time fluctuations, for example the model of Padmanabhan,\ \cite{pad}.
We conclude that the simple Poisson model for causal-set generation of
lambda fluctuations is constrained to produce an effectively never-present
lambda relative to the cold dark matter density in the Friedmann equation.

\textbf{Acknowledgement} I am grateful to Fay Dowker and Rafael Sorkin for
helpful discussions about the causal set model.

\end{document}